\documentstyle[twocolumn,twoside]{article}
\newcommand{\h}{\linebreak \hspace*{3ex}}
\newcommand{\hb}{\\ \hspace*{2ex}}

\begin{document}
\title{Cepheids as Distance Indicators: Some Current Problems}
\author{Michael Feast \\[2mm]
Astronomy Dept., University of Cape Town, Rondebosch, 7701, \hb 
 South Africa. {\em mwf@artemisia.ast.uct.ac.za}\\[2mm]
}
\date{}
\maketitle

ABSTRACT. A general review is given of the calibration of the
Cepheid distance scale, with particular reference to its use
in the determination of $\rm H_{0}$. Emphasis is placed on the
advantage of using a galactic calibration of the Cepheid scale,
rather than relying on an adopted distance to the LMC. It is
then possible to use LMC data to test for possible metallicity
effects on this scale.\\[1mm] 

{\bf 1. Introduction}\\[1mm]
Cepheid variables are at the present time of key importance for
establishing distances within our own Galaxy and to nearby galaxies
and as the basis for the determination of cosmological parameters.
As is well known, the determination of $\rm H_{0}$ by groups
working with the Hubble Space Telescope (e.g. Saha et al. 1999,
Freedman et al. 2001 = F2001) rests on Cepheid observations in relatively
nearby galaxies which are then used as calibrators of more
general distance indicators (e.g. SNIa, the Tully-Fisher relation,
surface brightness fluctuations etc.)
It is therefore important that every effort should be made to 
understand these stars and their luminosities, and particularly
to determine where present uncertainties lie.

The requirements of a primary distance indicator are easily stated.
They must be bright, so that they can be seen and measured at large
distances. They must also be recognizable for what they are. That is, 
they must not be confused with other types of objects. Their
absolute magnitudes must be accurately known. This means that
the intrinsic scatter in the absolute magnitudes must be small
and that these absolute magnitudes can be accurately calibrated
locally. There are two other considerations which are of equal
importance to these. Firstly, the reddenings must be measurable.
This is a crucial requirement. Secondly if there are effects due,
or likely to be due, to differing metallicities or ages between
the calibrators and the programme stars, these effects must
be known and measurable. In addition to all this, it is highly
desirable that the calibration and use of a primary distance
indicator should be as free as possible from theoretical
derivations or assumptions. Unless we can establish distance 
scales empirically we have no way of testing theory.

Very few objects, if any, can match the classical Cepheids in
fulfilling the requirements just set out. The main reason for this
is that they have been extensively studied both locally and
in the Magellanic Clouds for many years by a large number of
investigators. It can thus be fairly claimed that they are the
best understood of all pulsating variables. Other distance
indicators will be mentioned below, but it can also be claimed with
justification that Cepheids are the best understood of all the
distance indicators. 

The importance of Cepheids for distance
determinations rests of course on the period-luminosity (PL) relation.
The existence of this relation is well established,
particularly in the Magellanic Clouds. The PL relation has
relatively small scatter and the period-luminosity-colour
relation seems to have negligible intrinsic scatter. A detailed
discussion of some of these issues is given in Feast (1999).
Furthermore it will be shown below that a reliable 
calibration of the Cepheid scale can now be obtained locally.

Whilst therefore we can have some confidence in the Cepheids as
indicators of distance we should not be complacent. As will be
indicated later there are a number of areas where an improvement of
our knowledge of Cepheids is very desirable in order to  further
strengthen their use as distance indicators.\\[2mm]

{\bf 2. The HST Key Project Procedure}\\[1mm]
In order to focus the discussion, the present paper is limited to
a discussion of factors relevant to the use which is made
of Cepheids in the HST programmes for the determination of $\rm H_{0}$.
These depend on V and I photometry. 
In their ``final results" paper (F2001)
the HST Key Project workers adopt period-luminosity relations
at V and I  (PL(V) and PL(I)) of the following forms;
\begin{equation}
M_{V} = -2.760 \ \log \ P - 1.458,
\end{equation}
and
\begin{equation}
M_{I} = -2.962 \ \log \ P - 1.942.
\end{equation}
These relations are derived from the observations of LMC Cepheids
by Udalski et al. (1999, table 1), assuming the distance modulus
of the LMC is 18.50. For a Cepheid in a programme galaxy the
observed (intensity weighted) mean V and I are used with these
equations to derive apparent moduli. The difference between these
is $\rm E_{(V-I)}$ and multiplying this by 2.45 gives $\rm A_{V}$
and hence the true modulus.

The large number of Cepheids in the Udalski et al. work on the LMC
obviously makes this data-set attractive to use as a basis. It has, 
however, the drawback that individual reddenings of the Cepheids
themselves could not be determined (as can be done from multi-colour
photometry). The reddenings were derived by dividing the LMC into
a number of areas and estimating mean reddenings from the colours
of giant-branch-clump stars. The mean reddening of the Cepheids
used by Udalski et al. to derive equations 1 and 2 was
$\rm E_{(B-V)} = 0.147$ (not 0.10, as implied by F2001, section 3.1).
The method used by F2001 only requires the relative reddenings
of the LMC and their programme Cepheids. The absolute value of the
LMC reddening is not required. However, F2001 have not discussed
whether the Udalski et al. reddenings (which are considerably
larger than have been used in the past) are consistent with 
their adopted true modulus of the LMC. It would also be useful
in the future to have individual reddenings from multi-colour
photometry (see e.g. Caldwell and Coulson 1985) for many more
LMC Cepheids than are presently available. This could be used
to check whether equations 1 and 2 are affected by a dependence
of mean reddening on period
which would affect the slope of the
relations. In addition relations 1 and 2 depend
heavily on short period Cepheids ($\rm \log \ P \sim 0.5$)
(see Udalski et al. 1999 figs 2 and 3)
with very few at the long period end because of saturation
of the detector used by Udalski et al. A derivation of
the PL relation based on long period Cepheids would be important
since the weighted mean $\rm \log \ P$ of the F2001 programme Cepheids
is 1.42. In addition to this there is some evidence that Cepheid
moduli derived from V,I photometry require a metallicity dependent
correction. This will be discussed in detail below, where it
will be noted that the LMC Cepheids are metal-weak compared
with the mean of the HST Key Project sample which 
has a mean metallicity near solar.
Thus if Cepheid absolute magnitudes are based
on the LMC, a metallicity dependent uncertainty will enter.\\[2mm]

{\bf 3. A Galactic Calibration of Cepheids}\\[1mm]
A calibration of Cepheid luminosities and colours based on
observations in the general solar neighbourhood seems very
desirable. It would avoid basing distance scales on the metal-poor
LMC Cepheids. It would also be free of estimates of the LMC
distance based on non-Cepheid estimators, most of which are
less well understood than Cepheids themselves. However, as
discussed below these non-Cepheid LMC estimates can be used 
together with a Galactic Cepheid calibration, to place some
limits on metallicity effects on Cepheid luminosities. Such 
a Galactic Cepheid calibration is now possible.

Individual reddenings of many Galactic Cepheids have been derived
(Caldwell and Coulson 1987) based on the BVI system. These lead to
the relation;
\begin{equation}
<V>_{o} - <I>_{o} = 0.297 \ \log \ P + 0.427
\end{equation}
(see Feast 1999 where a slightly more exact procedure is given).
In this equation angle brackets denote intensity mean values.
Using a relation of this kind 
to obtain reddenings, together with a PL(V) relation is 
equivalent to the HST Key Project procedure.  

The PL(V) relation can be written;
\begin{equation}
M_{V} = \alpha \ \log \ P + \gamma.  
\end{equation}
Various estimates have been made for the slope $\alpha$
of this relation. Caldwell and Laney (1991) derived $-2.63 \pm 0.08$
from SMC Cepheids; The OGLE results for the LMC discussed in the previous
section gave $-2.76 \pm (0.03)$. The uncertainty is bracketed
since much of the weight is in very short period Cepheids.
Caldwell and Laney (1991) derived $-2.81 \pm 0.06$ for the LMC.
For Cepheids in the general solar neighbourhood Gieren et al.
(1998) obtained $-3.04 \pm 0.14$
from Baade-Wesselink luminosities. There is a slight hint here that
there may be a trend of slope with metallicity since this changes 
in the sequence, SMC, LMC, Galaxy. It would be valuable to study
this further, but at present there is no strong evidence of a 
significant trend of slope with metallicity. In the following
we adopt for the Galaxy the result of Caldwell and Laney
for the LMC, $-2.81 \pm 0.06$. This is the only result taken
from the LMC in the present calibration.

It is of interest to see how the Cepheid distance scale would be
affected by a change of slope from the adopted value, $-2.81$, 
to the Galactic value of Gieren et al., $-3.04$. The mean $\rm log \ P$ 
of the Cepheids used in F2001, weighted according to their
contributions to the final value of $\rm H_{0}$, is 1.42. 
Most of the Cepheids used as calibrators in the discussion below
are of shorter period. The weighted mean $\rm log \ P$ of the
parallax calibrators is 0.8 whilst for the clusters it is 1.1.
Changing the PL(V) slope from $-2.81$ to $-3.04$ would result in
an increase in the parallax scale when applied to the F2001 results
of 0.14 mag (a seven percent increase in distance scale) over
that actually adopted. 
Thus if a variation in slope with
metallicity in the sense tentatively suggested by the
possible SMC/LMC/Galaxy trend is actually confirmed
it would result in a decrease in the revised value of $\rm H_{0}$
discussed in section 7. 

A change of slope from the value adopted here (--2.81) to the
LMC OGLE value (--2.76), would result in a negligibly small
increase in $\rm H_{0}$.

There are four principal ways of obtaining a value of the PL(V)
zero-point $\gamma$ for galactic Cepheids (see Feast 2001).

1. A bias free analysis of the Hipparcos parallaxes of Cepheids
leads to $\gamma = -1.43 \pm 0.12$ (Feast and Catchpole 1997,
Feast 1999). This result and its bias free nature has been confirmed
both directly and with Monte Carlo simulations (Pont 1999,
Lanoix et al. 1999, Groenewegen and Oudmaijer 2000).

2. Hipparcos proper motions can be combined with radial velocities in
a statistical-parallax type solution. This requires a model. The 
dominant effect 
in the case of Cepheids is that of differential galactic rotation which is
clearly seen in the radial velocities and the proper motions
separately. The model therefore is rather firmly based. Using this 
method Feast, Pont and Whitelock (1998) obtained $\gamma = -1.47 \pm 0.13$.

One can attempt a solution using the solar motion obtained from
a combined discussion of solar motion and differential galactic
rotation using proper motions and radial velocities. In this way
the solar motion has a value which is averaged out over the
whole large region covered by the Hipparcos and radial velocity
Cepheids and is not confined to a small region near the Sun.
The results of Feast and Whitelock (1997) imply a scale which
is $0.04 \pm 0.26$ mag larger than that just given (Feast 2000). However, the
uncertainty is too large for this solution to make any 
significant contribution
to a final value. 

The above discussion refers to the use of the systematic motions
of the Cepheids.
Because the velocity dispersion of Cepheids is small,
any comparison of radial velocity and proper motion residuals will
be sensitive to the treatment of observational scatter and probably
also to group motions. 
 
3. Pulsation parallaxes (the Baade-Wesselink method in its
various forms) requires a number of assumptions to be made regarding
such things as limb darkening, the colour-surface brightness relation
etc. Thus whilst the internal consistency of the method is good,
the external uncertainty is difficult to estimate. Feast (1999)
derived, $\gamma = -1.32 \pm 0.04$(internal) from the discussion of
Laney (1998). The angular diameters of a few Cepheids have
recently been determined interferometrically (Kervella et al. 2001;
Nordgren et al. 2000; Armstrong et al. 2001) and 
the change in angular diameter of $\zeta$ Gem due to pulsation has
been detected (Lane et al. 2000). When more measurements along these lines
are made it should be possible to refine the pulsation parallax
method further.

4. Young open clusters containing Cepheids can be used to derive
$\gamma$ provided the cluster scale is known. Feast (2001) shows how
this scale can be rather firmly based on the well-determined
Hipparcos parallax of the Hyades. This leads to 
$\gamma = -1.43 \pm 0.05$(internal).

Because some of the error estimates are internal only, it is not
safe to use these errors to weight the four estimates
of $\gamma$ given above. A straight mean gives $-1.41$. The uncertainty
in this value is probably somewhat less than 0.10.
It should be noted that methods 1 and 2 for determining $\gamma$
lead to distance scales which do not depend on the zero-point
of the reddening scale adopted, so long as consistent reddenings
are applied to both calibrating and programme Cepheids. This is
not the case for the other two methods. This is particularly so
in the case of the clusters where the main sequence fitting
depends sensitively on the adopted reddening in a way which does
not cancel out in application to programme Cepheids.\\[2mm]

{\bf 4. Tests for Metallicity Effects. I}\\[1mm]
It has long been believed 
from observations of metal-poor Cepheids in the Magellanic Clouds
that there is a metallicity effect at least
in the (B--V) colours of Cepheids. Laney (quoted by Feast 1991)
showed that the BVI colours of SMC Cepheids could not be brought
into agreement with those in the LMC unless such an effect was assumed
to exist due to the SMC Cepheids being more metal-poor than those in the 
LMC (a result which is known spectroscopically). 

There are at least
three effects attributable to metallicity differences which could
affect the PL relation at a given wavelength.

1. Laney and Stobie (1986) showed from infrared photometry 
of galactic and Magellanic Cloud Cepheids that metallicity changes
lead to a change in surface temperature at a given pulsation period.

2. There must be some effect (especially at the shorter wavelengths)
due to a change of blanketing with metallicity at a given surface
temperature.

Both the above have the effect of changing the bolometric
corrections applicable at different wavelengths.

3. Although it is generally assumed that the bolometric PL
relation is insensitive to metallicity changes, not all theorists
agree on this issue.

Laney (1999) showed that the radii of Magellanic Cloud Cepheids
as derived from Baade-Wesselink analyses fitted the galactic
period-radius relation. Infrared photometry (Laney and Stobie 1986)
shows that at a given period the metal-poor Cepheids in the Clouds
are slightly hotter than the galactic ones. The evidence thus suggests
that the bolometric luminosities of Cepheids of a given period increases
with decreasing metallicity. However, the effect is small and within
the uncertainties of the measurements.

Evidently in the case of the HST work we require to estimate the
effects of metallicity  changes on equations 3 and 4 above 
(or equivalently, equations 1 and 2).
There has been some confusion in the literature since the effect
on the derived distance moduli due to a metallicity change
is found to be in opposite directions for the equations 3 and 4. It 
is thus essential to discuss the combined effect on these
two equations of a change in metallicity.

The most direct empirical test of the effect of metallicity
in determining distance moduli from V,I photometry is 
that carried out by Kennicutt et al. (1998). They observed
Cepheids in the galaxy M101 at different distances from the centre.
At these different positions they could estimate abundances
([O/H]) from HII region observations. Their results lead to
 a metallicity effect on Cepheid distance modulus determinations of 
$\rm 0.24 \pm 0.16 \ [O/H]^{-1}$
in the sense that without the correction
the distance of a metal-poor Cepheid would
be overestimated.
This result suggests that a
metallicity effect in the V,I method exists, but the uncertainty
is evidently still large.
(For a more detailed discussion of metallicity effects, see
Feast (1999)).\\[2mm]

{\bf 5. Tests for Metallicity Effects. II. The LMC.}\\[1mm]
It has often been suggested that the Cepheid distance scale can
(and should) be determined by deriving the distance to the
LMC in some non-Cepheid way
and then using this as the standard. However, quite apart from possible
metallicity effects on the Cepheid scale, it must be born in mind
that all non-Cepheid distance indicators so far used to
estimate the distance of the LMC have their own problems and
uncertainties. These problems include the calibration of
these other indicators, their possible metallicity dependence,
and, their reddenings relative to the Cepheids. Nevertheless,
these non-Cepheid indicators can give a useful indication of
a probable metallicity effect on Cepheid moduli. 

Non-Cepheid moduli of the LMC were discussed by Feast (2001). The
following is a slightly updated summary of that discussion
which should be consulted for details. 

1. RR Lyraes. The absolute magnitudes
of RR Lyrae stars can be derived from; parallaxes
(Koen and Laney 1998), parallaxes of
horizontal branch stars
(Gratton 1998), globular clusters with distances derived
from sub-dwarf fitting
(Carretta et al. 2000), $\delta$ Sct parallaxes 
(McNamara 1997) and statistical
parallaxes
(Gould and Popowski 1998). These can be combined with the data on LMC
field RR Lyraes (Clementini et al. 2000) to obtain an LMC true
distance modulus of 18.54.

2. Mira Variables. The infrared (K) period-luminosity relation
for Miras can be calibrated using Hipparcos parallaxes (Whitelock
and Feast 2000) and also using Miras in globular clusters with cluster
distances on the subdwarf scale of Carretta et al. (2000). Using these
calibrations with the LMC Mira data (Feast et al. 1989) gives
(Feast, Whitelock and Menzies, to be published) an LMC modulus
of 18.60.

3. The ring round SN1987A. The best estimate of the LMC modulus
from this is probably that of Panagia (1998) which is 18.58.

4. LMC globular clusters. Main-sequence fitting (Johnson et al. 1999)
of LMC clusters can be used to derive a modulus of 18.52.

5. The red giant clump. The use of the red giant clump as an LMC
distance indicator is complicated by age and metallicity effects
and by reddening uncertainties. The best estimate is probably
that of Girardi and Salaris (2001) who find a modulus of 18.55.

6. Eclipsing Variables. Although the use of eclipsing variables as
distance indicators seems rather straight forward in theory, its
application to the LMC requires at present a number of assumptions.
The best current estimated 
of the modulus by this method is probably that of
Groenwegen and Salaris (2001) who obtain 18.42.

The real uncertainties of the above estimates are probably about
0.1mag. The first three estimates seem likely to be the most secure
and a straight mean of them is 18.57. A straight mean of all six
estimates is 18.54. The V,I distance modulus of the LMC using
equations 3 and 4 and $\gamma = -1.41$ is 18.66, uncorrected
for metallicity effects. If we adopt a metallicity effect
of $\rm 0.2 \ mag \ [O/H]^{-1}$ and $\rm [O/H]_{LMC} = -0.4$ as
is done by F2001 following Kennicutt et al. (1998),
we obtain a corrected Cepheid modulus of 18.58. The close
agreement of this value with the mean non-Cepheid estimate is no
doubt partly fortuitous since the uncertainties in both estimates
are probably of order 0.1. However, the results do suggest that a
metallicity correction of the approximate amount suggested by
Kennicutt et al. is present in V,I estimates.\\[2mm]

{\bf 6. A Cepheid test using NGC4258}\\[1mm] 
  A distance to the galaxy NGC4258 has been derived from the motion
of $\rm H_{2}O$ masers in the central region 
and a simple model (Herrnstein et al. 1999). Newman et al. (2001)
have recently published HST V,I observations of Cepheids in this galaxy
which can thus also be used to derive a distance. The metallicity
adopted by Newman et al. (from HII region measurements by Zaritsky et al.
1994) is slightly below solar ($\rm [O/H] = -0.05$). Thus any reasonable
metallicity correction to a galactic calibration will be very small.
Adopting the galactic calibration given above and a metallicity effect
of $\rm 0.20 \ mag \ [O/H]^{-1}$ one obtains a true distance modulus
of $29.53 \pm 0.17$, where the standard error is taken from the
discussion of Newman et al. together with an estimated error of the
galactic zero-point of $\sim 0.10$. In deriving this value we have followed
the procedure of Newman et al. and used template-fitted DoPHOT 
mean magnitudes kindly supplied by Dr Newman. These magnitudes differ
slightly from the values given in Table 2 of Newman et al.
A metallicity correction of $-0.01$ mag has been applied. 
The Cepheid distance is therefore greater than the maser one by
$0.24 \pm 0.21$ mag. This difference is not significant.\\[2mm]

{\bf 7. Key Project Value of $\rm H_{0}$ based on a Galactic Calibration}\\
[1mm] 
F2001 have summarized the HST Cepheid Key Project
data together with related HST data by other groups (especially
the SNIa group). As indicated in the Introduction they use an
adopted LMC modulus as the
basis of their analysis. They also introduce a metallicity
correction for the first time in their series of papers.
They derive a value of 
$\rm H_{0} = 72 \pm 8 \ km \ s^{-1} \ Mpc^{-1}$.  
Using their data and with the same metallicity term 
($\rm 0.2 \ mag \ [O/H]^{-1}$) but with the galactic calibration
derived above, one obtains $\rm H_{0} = 67 \pm 8$ where the standard
error is taken from F2001. These two estimates of $\rm H_{o}$
are gratifyingly close. Amongst the reasons for preferring 
a scale based on the galactic calibration is that it is practically
immune to the uncertain metallicity correction. This follows since
the mean metallicity of
the F2001 Cepheid galaxies, weighted according to their contribution
to the final value of $\rm H_{0}$ is close to solar ([O/H] = --0.08).
It has been hypothesized that the metallicity correction could be as
high as $\rm 0.6 \ mag \ [O/H]^{-1}$. Whilst it seems unlikely that it could
be as large as this, even such a large value will have only a very small
effect on $\rm H_{0}$ derived using the galactic calibration. On
the other hand, since $\rm [O/H]_{LMC} = -0.4$, such a large correction
coefficient
would have a significant effect (a six percent decrease in $\rm H_{0}$)
on a calibration based on an adopted LMC modulus.

It should be made clear that the above discussion is given to illustrate
how the galactic calibration affects the conclusions of F2001.
There appears still to be considerable differences in the interpretation
of the HST data by different groups, unrelated to the adopted basic
Cepheid distance scale (see for instance Saha et al. 1999).  
Full agreement on these matter is required before the value of
$\rm H_{0}$ can be considered properly established.\\[2mm] 

{\bf 8. Conclusions}\\[1mm]
The calibration of the Cepheid zero-point is now quite well established
from galactic observations. The present uncertainty is about 0.10 mag.
It is obviously desirable to improve this accuracy, though it is
not clear that this can be achieved without further astrometry
from space (GAIA etc.). It is evident that further work is needed
on the PL(V) and PL(V--I) slopes at long periods and it would
be very desirable to determine empirically if these slopes  depend
on metallicity.
In doing this it would be essential to derive individual reddenings
for the Cepheids from multicolour photometry.
It would also be of considerable interest to 
measure the metallicities of many more galactic Cepheids. In
particular it would be useful to know more precisely the spread
in metallicities amongst local Cepheids.\\[2mm]

{\bf Acknowledgements.}\\[1mm] 
 I am grateful to Patricia Whitelock and
John Menzies for allowing me to refer to work in progress.\\[3mm] 

\indent
{\bf References}\\[2mm] 
Armstrong J.T. et al.: 2001, {\it Astron.\ J.}, {\bf 121}, 476.\\
Caldwell J.A.R., Coulson I.M.: 1985, {\it Mon.\ Not.\ R.\ \h Ast. Soc.},
{\bf 212}, 879, (erratum {\bf 214}, 639).\\
Caldwell J.A.R., Coulson I.M.: 1987, {\it Astron.\ J.}, {\bf 93}, \h 1090.\\
Caldwell J.A.R., Laney C.D.: 1991, in {\it The Magellanic \h Clouds}, Proc.
IAU Symp.\ 148, 249.\\
Carretta E., Gratton R.G., Clementini G., Fusi Pecci \h F.: 2000,
{\it Ap.\ J}, {\bf 533}, 215.\\
Clementini G., Gratton R., Bragaglia A., Carretta E., \h Di Fabrizio L.:
2000, astro-ph/0007471.\\
Feast M.W.: 1991, in {\it Observational Tests of Cosmo- \h logical Inflation},
Kluwer, Dordrecht, p.147.\\ 
Feast M.W.: 1999, {\it Pub.\ Ast.\ Soc.\ Pacif.}, {\bf 111}, 775.\\
Feast M.W.: 2000, {\it Mon.\ Not.\ R.\ Ast.\ Soc.}, {\bf 313}, 596.\\ 
Feast M.W.: 2001, in {\it New Cosmological Data and \h the Values
of the Fundamental Parameters}, Proc.\ \h IAU Symp.\ 201, 17.\\
Feast M.W., Catchpole R. M.: 1997, {\it Mon.\ Not.\ R.\ Ast.\ \h Soc.},
{\bf 286}, L1.\\ 
Feast M.W., Glass I.S., Whitelock P.A., Catchpole \h R.M.: 1989,
{\it Mon.\ Not.\ R.\ Ast.\ Soc.}, {\bf 241}, 375.\\
Feast M.W., Pont F., Whitelock P.A.: 1998, {\it Mon.\ Not.\ \h R.\ Ast.\
Soc.}, {\bf 298}, L43.\\
Feast M.W., Whitelock P.A.: 1997, {\it Mon.\ Not.\ R.\ Ast.\ \h Soc.},
{\bf 291}, 683.\\
Freedman W.L. et al.: 2001, {\it Ap.\ J.}, {\bf 553}, 47. (F2001)\\
Gieren W.P, Fouqu\'{e} P., G\'{o}mez M.: 1998, {\it Ap.\ J.}, \h 
{\bf 496}, 17.\\
Giradi L., Salaris M.: 2001, {\it Mon.\ Not.\ R.\ Ast.\ Soc.}, \h 
{\bf 323}, 109.\\
Gould A., Popowski P.: 1998, {\it Ap.\ J.}, {\bf 508}, 844.\\
Gratton R.G.: 1998, {\it Mon.\ Not.\ R.\ Ast.\ Soc.}, {\bf 296}, 739.\\
Groenewegen M.A.T., Oudmaijer R.D.: 2000, {\it Astron.\ \h Astrophys.},
{\bf 356}, 849.\\
Groenewegen M.A.T., Salaris M.: 2001, {\it Astron.\ Astro- \h phys.},
{\bf 366}, 752.\\
Herrnstein J.R. et al.: 1999, {\it Nature}, {\bf 400}, 539.\\ 
Johnson J.A., Bolte M., Stetson P.B., Hesser J.E., \h Somerville R.S.:
1999, {\it Ap.\ J.}, {\bf 527}, 199.\\
Kennicutt R.C. et al.: 1998, {\it Ap.\ J.}, {\bf 498}, 181.\\
Kervella P. et al.: 2001, {\it Astron.\ Astrophys.}, {\bf 367}, 876.\\
Koen C., Laney.: 1998, {\it Mon.\ Not.\ R.\ Ast.\ Soc.}, {\bf 301}, \h 582.\\
Lane B.F., Kuchner M.J., Boden A.F., Creech-Eakman \h M.,
Kulkarni S.R.: 2000, {\it Nature}, {\bf 407}, 485.\\
Laney C.D.: 1998 in {\it A Half-Century of Stellar Pulsa- \h tion
Interpretations}, ASP.\ Conf.\ Ser., {\bf 135}, 180.\\
Laney C.D.: 1999 in {\it The Stellar Content of Local \h Group Galaxies},
IAU Symp.\ 192, 459.\\
Laney C.D., Stobie.: 1986, {\it Mon.\ Not.\ R.\ Ast.\ Soc.}, \h  
{\bf 222}, 449.\\
Lanoix P., Paturel G., Garnier R.: 1999, {\it Mon.\ Not.\ R.\ \h Ast.\
Soc.}, {\bf 308}, 969.\\
McNamara D.H.: 1997, {\it Pub.\ Ast.\ Soc.\ Pacif.}, {\bf 109}, \h 1232.\\
Newman J.A. et al.: 2001, {\it Ap.\ J.}, {\bf 553}, 562.\\
Nordgren T.E. et al.: 2000, {\it Ap.\ J.}, {\bf 543}, 972.\\
Panagia N.: 1998, {\it Mem.\ Soc.\ Astron.\ Italiana}, {\bf 69}, 225.\\
Pont F.: 1999 in {\it Harmonizing the Cosmic Distance \h Scale in the
Post-Hipparcos Era}, ASP.\ Conf.\ Ser., \h {\bf 167}, 113.\\ 
Saha A. et al.: 1999, {\it Ap.\ J.}, {\bf 522}, 802.\\
Udalski A., et al.: 1999, {\it Acta Astr.}, {\bf 49}, 201.\\
Whitelock P.A, Feast M.W.: 2000, {\it Mon.\ Not.\ R.\ Ast.\ \h Soc.},
{\bf 319}, 759.\\
Zaritsky D., Kennicutt R.C., Huchra J.P.: 1994, {\it Ap.\ \h J.},
{\bf 420}, 87.\\
\vfill
\end{document}